\documentclass[prb,twocolumn,preprintnumbers,amsmath,amssymb,superscriptaddress]{revtex4}
\usepackage[paperwidth=210mm,paperheight=297mm,centering,hmargin=2.0cm,vmargin=2.6cm]{geometry}
\usepackage{graphicx}
\usepackage{dcolumn}
\usepackage{bm}
\usepackage{upgreek}
\usepackage{amsmath}
\usepackage{url}
\usepackage{hyperref}
\usepackage{lipsum}
\usepackage{inputenc}

\newcommand{\beq}{\begin{equation}}
\newcommand{\eeq}{\end{equation}}
\newcommand{\bea}{\begin{align}}
\newcommand{\eea}{\end{align}}
\newcommand{\beqa}{\begin{eqnarray}}
\newcommand{\eeqa}{\end{eqnarray}}

\begin{document}

\title{Opto-Mechanical Tuning of the Polarization Properties of Micropillar Cavity Systems with embedded Quantum Dots}

\author{Stefan Gerhardt}
\affiliation{Technische Physik, Physikalisches Institut and W\"urzburg-Dresden Cluster of Excellence ct.qmat, Universit\"at W\"urzburg, Am Hubland, D-97074 W\"urzburg, Germany}
\author{Magdalena Mocza{\l}a-Dusanowska}
\affiliation{Technische Physik, Physikalisches Institut and W\"urzburg-Dresden Cluster of Excellence ct.qmat, Universit\"at W\"urzburg, Am Hubland, D-97074 W\"urzburg, Germany}
\author{{\L}ukasz Dusanowski}
\affiliation{Technische Physik, Physikalisches Institut and W\"urzburg-Dresden Cluster of Excellence ct.qmat, Universit\"at W\"urzburg, Am Hubland, D-97074 W\"urzburg, Germany}
\author{Tobias Huber}
\affiliation{Technische Physik, Physikalisches Institut and W\"urzburg-Dresden Cluster of Excellence ct.qmat, Universit\"at W\"urzburg, Am Hubland, D-97074 W\"urzburg, Germany}
\author{Simon Betzold}
\affiliation{Technische Physik, Physikalisches Institut and W\"urzburg-Dresden Cluster of Excellence ct.qmat, Universit\"at W\"urzburg, Am Hubland, D-97074 W\"urzburg, Germany}
\author{Javier Mart\'in-S\'anchez}
\affiliation{Department of Physics, University of Oviedo, Oviedo, Spain}
\affiliation{Center of Research on Nanomaterials and Nanotechnology, CINN (CSIC-Universidad de Oviedo), El Entrego 33940, Spain}
\author{Rinaldo Trotta}
\affiliation{Department of Physics, Sapienza University of Rome, Piazzale A. Moro 5, 00185 Rome, Italy}
\author{Ana Predojevi\'{c}}
\affiliation{Department of Physics, Stockholm University, SE-106 91 Stockholm, Sweden}
\author{Sven H\"ofling}
\affiliation{Technische Physik, Physikalisches Institut and W\"urzburg-Dresden Cluster of Excellence ct.qmat, Universit\"at W\"urzburg, Am Hubland, D-97074 W\"urzburg, Germany}
\affiliation{SUPA, School of Physics and Astronomy, University of St Andrews, St Andrews, KY16 9SS, United Kingdom}
\author{Christian Schneider}
\affiliation{Technische Physik, Physikalisches Institut and W\"urzburg-Dresden Cluster of Excellence ct.qmat, Universit\"at W\"urzburg, Am Hubland, D-97074 W\"urzburg, Germany}

\date{\today}

\begin{abstract}

Strain tuning emerged as an appealing tool to tune fundamental optical properties of solid state quantum emitters. In particular, the wavelength and fine structure of quantum dot states could be tuned using hybrid semiconductor-piezoelectric devices. Here, we show how an applied external stress can directly impact the polarization properties of coupled InAs quantum dot-micropillar cavity systems. In our experiment, we find that we can reversibly tune the anisotropic polarization splitting of the fundamental microcavity mode by approximately 60 $\upmu$eV. We discuss the origin of this tuning mechanism, which arises from an interplay between elastic deformation and the photoelastic effect in our micropillar. Finally, we exploit this effect to tune the quantum dot polarization opto-mechanically via the polarization-anisotropic Purcell effect. Our work paves the way for optomechanical and reversible tuning of the polarization and spin properties of light-matter coupled solid state systems. 
\end{abstract}
\maketitle

\section{Introduction}

Micropillar cavities are one of the widely used design implementations of high-performance, solid state single photon sources \cite{Kok2007, Brien2007, Gerard1998, Gerard2001, Santori2001, Santori2002, Pelton2002Efficient, Senellart2017}, microlasers operating in the weak \cite{Stock2013, Munnelly2015} and strong coupling regime \cite{Bajoni2008Polariton, Klaas2018photon-number}, and non-linear photonic crystal lattices \cite{Pinto_Amo2012, Klembt2017}. The behaviour of a quantum dot (QD) embedded in such a cavity is described by cavity quantum electrodynamics \cite{Gerard1998, Gerard2001}. In particular, by making use of the Purcell effect, it is possible to significantly improve the QD performance \cite{Vuvckovic2003enhanced, Santori2004}, enabling efficient collection of single photons with near unity indistinguishability \cite{Uns16_insitu, Ding2016, somaschi2016near}. Deterministic fabrication of such micropillar devices yields great improvements in the spatial and spectral alignment of the cavity resonance and the QD emission \cite{dousse2008, He2017}. Nevertheless, spectral fine-tuning remains the missing tool required to overcome remaining fabrication inaccuracies. Temperature and electrical tuning techniques cause a significant deterioration of the source performance via phonon-induced decoherence \cite{Gerhardt18} and carrier tunneling, respectively \cite{Hermannstadter2010Influence}. Therefore, strain tuning techniques were developed allowing to reversibly shift emitter energies without degrading their optical properties \cite{Martin-Sanchez2017}. Recently, these techniques were also implemented to tune the QD emitters coupled to micropillar cavities by applying mechanical stress \cite{Moczala2019Strain}. Here, we report on tuning of the polarization of the cavity's fundamental optical mode by anisotropic strain, and discuss how the extrinsic stress impacts the photonic resonance of the micropillar. This new tuning mechanism directly enables us to shape the polarization of a QD in the weak cavity coupling regime, taking advantage of the Purcell effect. Finally, we provide insights into the physics of our mechanically tunable light-matter coupled system, and propose a variety of possible applications achievable with our platform. \\


\begin{figure*}[htb]
\begin{center}
\includegraphics[width=1.00\textwidth]{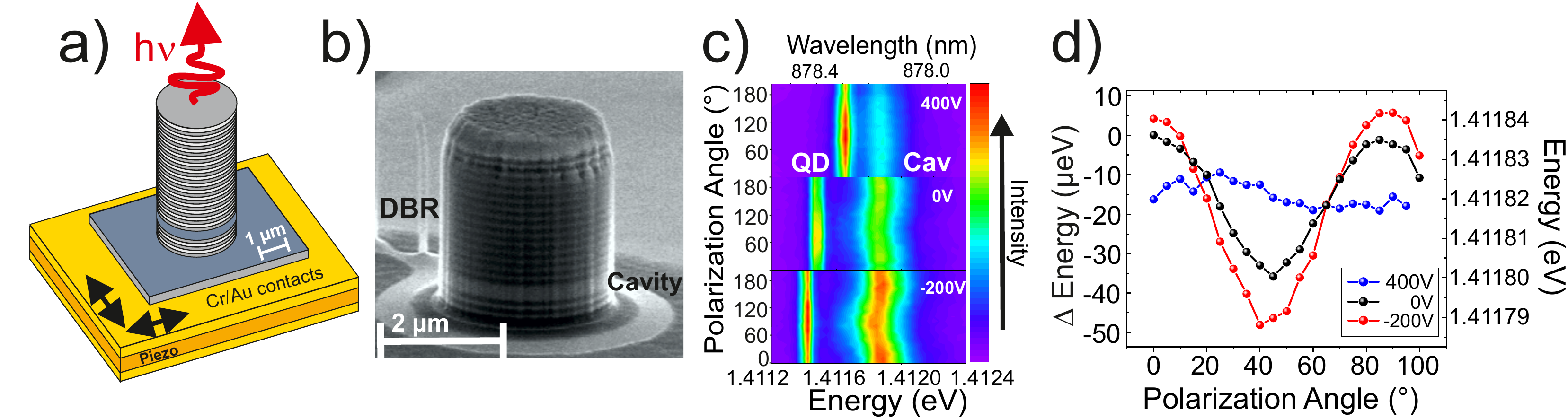}
\caption{(a) Illustration of a QD-micropillar cavity (grey colours) on a Cr/Au-coated piezoelectric substrate (yellow). (b) SEM image of the investigated micropillar cavity  with a diameter of d = 2.8~$\upmu$m before planarization. (c) Polarization resolved photoluminescence spectra. With applied stress the emitter shifts its energy and the linear polarization splitting of the cavity resonance is modified. (d) Magnitude of the linear polarization splitting of the fundamental cavity mode for different applied voltages. The circular symmetry of the pillar, which features a H/V splitting of 35~$\upmu$eV without applied strain, can be almost fully restored by applying a voltage of 400~V. By decreasing the applied bias voltage to -200~V, the splitting is enhanced to $\mathrm{\Delta E_{H,V}}$ = (52.4~$\pm$~0.4) ~$\upmu$eV.}
\label{fig:Pillar_Piezo_Contour_Splitting}
\end{center}
\end{figure*}

We studied a sample based on an AlAs/GaAs microcavity structure with embedded expitaxially grown self-assembled InGaAs QDs as the active medium. The optical confinement in the growth direction was enabled by two stacks of 15 and 25 AlAs/GaAs mirror pairs forming the upper and lower distributed Bragg reflector (DBR), respectively. In a first step, the GaAs substrate was mechanically lapped down to a thickness of approximately 30~$\upmu$m. By means of an epoxy-based photoresist (SU8) this planar sample was bonded onto a 300~$\upmu$m thick (001)[Pb(Mg$_{1/3}$NB$_{2/3}$)O$_3$]$_{0.72}$[PbTiO$_{3}$]$_{0.28}$ (PMN-PT) piezoelectric substrate, which was coated with chromium$/$gold contacts \cite{Trotta2012Nanomembrane, Trotta2016}. Via high resolution electron beam lithography and a subsequent lift-off process, micropillars were defined on the planar sample and transferred into the heterostructure via reactive ion etching (RIE) (Ar/Cl$_2$ plasma). To guarantee an adequate strain transfer to the QD-micropillar system, only two bottom DBR mirror pairs were etched. As a final step, the sample was planarized with benzocyclobutene (BCB) polymer to mechanically stabilize the micropillars and protect the sidewalls from oxidation. The final device is illustrated in Fig.~\ref{fig:Pillar_Piezo_Contour_Splitting}(a) \cite{Moczala2019Strain}.

To ensure heat transfer and enable electrical contacts to the piezoelectric actuator via wire bonding, the device was mounted onto an AlN chip carrier. Fig.~\ref{fig:Pillar_Piezo_Contour_Splitting}(b) depicts a scanning electron microscope (SEM) image of a micropillar with a diameter of 2.8~$\upmu$m and a height of approximately 3~$\upmu$m.\\


We investigated the impact of applied stress on the polarization properties and the structure of the fundamental cavity resonance. In particular anisotropic strain, supplied by the piezoelectric actuator, is expected to influence both the crystal structure (and thus the material's bandgap via the deformation potentials) as well as the geometry of the circular micropillar. To test this effect, we selected a device which had a QD red-shifted with respect to the cavity mode. The QD emission showed no distinct linear polarization features. Hence, we attributed it to a trion state. Fig.~\ref{fig:Pillar_Piezo_Contour_Splitting}(c) shows three series of polarization-resolved photoluminescence spectra, which were recorded with different bias applied to the piezoelectric crystal. Evidently, the QD emission experiences a spectral shift due to the modification of the confined states, and it approaches the cavity resonance with increasing positive voltage, as discussed in detail in our previous work \cite{Moczala2019Strain}. In order to gain access to the polarization properties of the device, we investigated the photoluminescence as a function of the linear polarization angle. The peak energy of the luminescence spectrum of the cavity experiences an oscillatory behaviour at 0~V as the linear polarization axis in the detection is varied. This oscillation becomes more pronounced when a negative bias of -200~V is applied to the actuator. However, it notably reduces for a positive bias of 400~V. This oscillation is a result of detecting two orthogonally, linear polarized resonances split by less than a linewidth. The position of the centre of the peak as a function of the polarization angle is shown in Fig.~\ref{fig:Pillar_Piezo_Contour_Splitting}(d). We achieved an overall tuning range of the H/V cavity mode splitting of $\mathrm{\Delta E_{H,V}}$ = (52.4~$\pm$~0.4)~$\upmu$eV with applied negative bias, and restored the polarization degeneracy by applying positive bias. In contrast to the quantum dot emission, the mean energy of the cavity resonance stays fully unaffected.

\begin{figure}[htb]
\begin{center}
\includegraphics[width=0.3\textwidth]{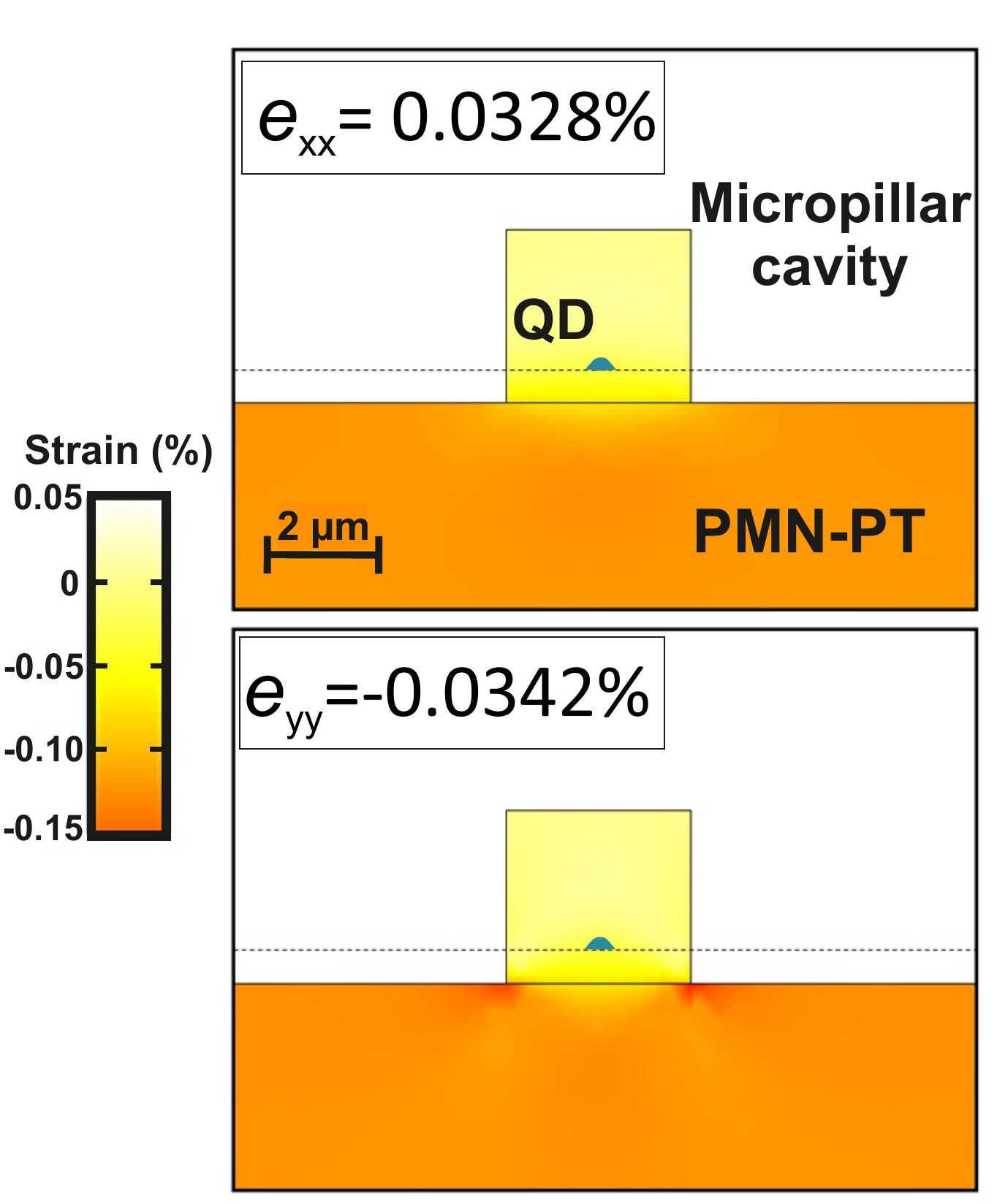}
\caption{Strain maps on micropillars embedding QDs obtained by FEM simulations with -200~V applied to the actuator. The numbers refer to the strain at the QD position. The case of nearly isotropic strain fields by employing a (001) PMN-PT piezoelectric plate is shown. A small in-plane anisotropy in the strain could be simulated.}
\label{fig:stress}
\end{center}
\end{figure}

The tuning of $\mathrm{\Delta E_{H,V}}$ via external strain has two possible origins. First, straining the sample can yield a shape anisotropy, and thus modify the ellipticity of our device. Second, the mechanic deformation changes the refractive index of the micropillar along the two main axis as a consequence of the photoelastic effect \cite{Kirkby1979photoelastic}.
The photoelastic tensor directly connects the elastic deformation of the cylindrical structure and the dielectric constant. Here, we consider the transversal electric (TE) wave in our pillar. As detailed in Kirkby et al. \cite{Kirkby1979photoelastic}, the impact of strain and stress to the dielectric constant in GaAs can be expressed via

\begin{align}
\Delta \epsilon_{r_{xx}}= -\epsilon_r^2 \left(\Delta e_{xx}\left[\frac{1}{2}(p_{11}+p_{12})+p_{44}\right]+\Delta e_{zz}p_{12}\right),
\label{eq:ela}
\end{align}

with the photoelastic coefficients $p_{11}$ = -0.165, $p_{12}$ = -0.140, $p_{44}$ = -0.072 given by Dixon \cite{Dixon1967Photoelastic} and $\epsilon_r$ being the dielectric constant. Our simulations show that $\mathrm{\Delta e_{zz}}$ is negligible as compared to $\Delta e_{xx}$ (even if the inplane strain is anisotropic). Therefore, we set it equal to 0. We can estimate the change of the mode splitting of a moderat elliptical pillar, based on the two above given interconnected phenomena. To do so, first, we estimated the extent of the mode splitting caused by the pillar ellipticity \cite{Reitzenstein2007}. Subsequently, we derived an equivalent term to gain a quantitative value for the change of the dielectric constant \cite{Gutbrod1998}. This term results in

\begin{align}
\Delta E(\Delta r_{c}, \Delta \epsilon_{r}) = \frac{\hbar^2 c^2 \chi_{0,1}^2}{E_{circ}} \left(\frac{1}{r_c^3 \epsilon_r}  \Delta r_c + \frac{1}{2 r_c^2 \epsilon_r^2} \Delta \epsilon_r \right),
\label{eq:splitting}
\end{align}

with $E_{circ}$ being the emission energy of a circular micropillar with radius $r_c$ and $\chi_{0,1}^2$ denoting the first zero of the Bessel function $\mathrm{J_{n_{\phi}}(x_{n_{\phi},n_r} r/r_c)}$. 

By solving eq. \ref{eq:ela} and \ref{eq:splitting} simulteanously for our experimentally extracted $\mathrm{\Delta E_{H,V}}$ of $(52.4~\pm~0.4) ~\upmu$eV, we can estimate the contributions of the shape anisotropy and the photoelastic effect yielding $\Delta r_c$ = 4.8~nm and $\Delta \epsilon_r$ = 0.13. Therefore we can attribute a contribution of 40.5~\% to the splitting by the small ellipticity and of 59.5~\% by the change of the refractive index. \\

In order to provide a better understanding of the observed phenomenon, we have performed finite-element-method~(FEM) simulations that estimate the overall amount of strain induced on the QD-micropillar system. The simulations were obtained using the software Comsol Multiphysics and the piezoelectric constants provided by the company supplying the piezo material. Fig.~$\ref{fig:stress}$ shows the $e_{xx}$ and $e_{yy}$ components of the system's strain tensor when a bias of -200~V is applied to a (001) PMN-PT piezoelectric plate.\newline
As a consequence of the device geometry, a considerable strain relaxation occurs across the pillar and only about 20~\% of the strain provided by the piezoelectric actuator is transferred to the overlying QD structure. This explains the etching procedure we used where we do not to etch through the entire pillar but rather stop at the first two DBR pairs in the bottom segment. The overall hydrostatic strain $\mathrm{(e_{xx} + e_{yy})}$ is also quantitatively consistent with the blue shift of the QD emission lines and matches the values reported in previous works \cite{Trotta2015PRL, Trotta2012Nanomembrane}.

However, the simulations alone are not sufficient to explain why the cavity mode stays constant while the mode splitting changes when the bias is varied. We believe that the preserved energy of the cavity mode arises from the interplay between the dimension of the cavity and the change of the refractive index. However, the change of the mode splitting suggests that the strain delivered by the piezo is not completely isotropic in the plane, since the extremely small anisotropy $\mathrm{\epsilon~=~(\textit{e}_{xx}-\textit{e}_{yy})}$ causes a change of each axis of around 1~nm and cannot explain the voltage induced splitting observed in the experiment. The existence of this anisotropy is indeed consistent with previous findings \cite{Kumar2014} and it is most likely related to imperfections arising in the wafer-bonding process \cite{Ziss2017comparison}. We note that in this scenario, a highly anisotropic strain with $\mathrm{e_{xx}= -0.1~\%}$ and $\mathrm{e_{yy}=0.24~\%}$ is necessary to yield a tuning range as observed in the experiment.\\

\begin{figure*}[htb]
\begin{center}
\includegraphics[width=0.96\textwidth]{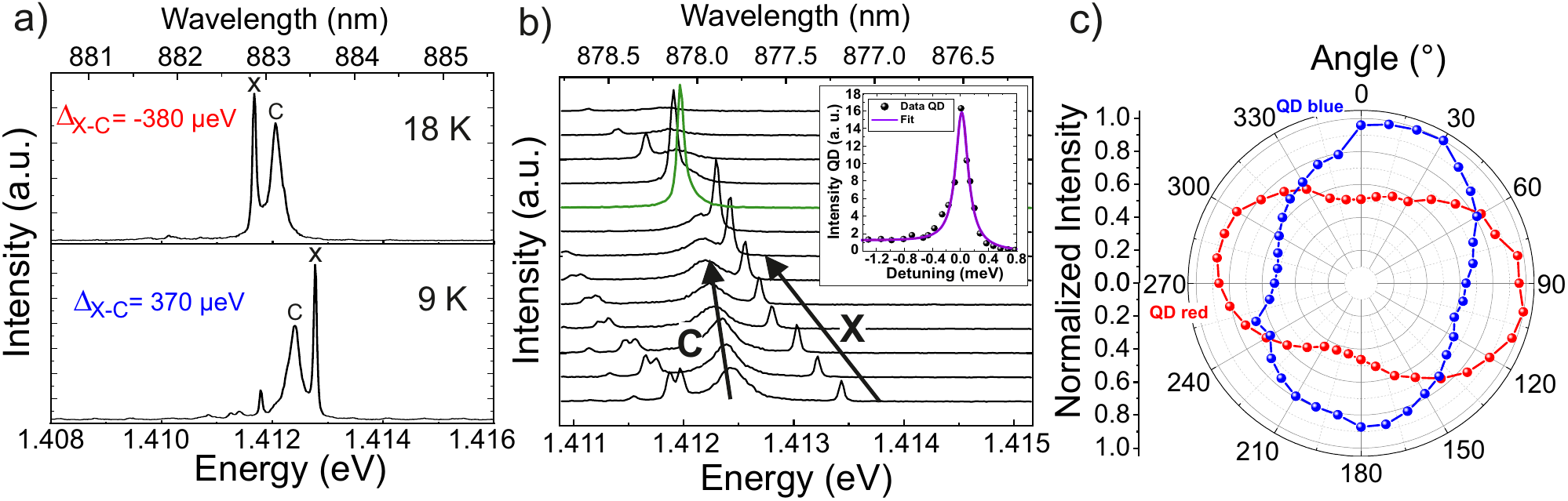}
\caption{(a) Spectra of the coupled QD-cavity system at two sample temperatures. The QD is red detuned at 18~K and blue detuned at 9~K. (b) Waterfall plot of spectra of QD (X) and cavity (C) with different detuning, resonance conditions labeled in green. Inset: The intensity increase of the QD emission indicates a maximum Purcell factor of $\mathrm{F_P}$ = 3.1\,$\pm$\,0.4. The polar plot in (c) reveals the perpendicular polarization orientation of the QD for red- and blue detuning conditions.}
\label{fig:contour_pol_purcell}
\end{center}
\end{figure*}

The tuning of the cavity polarization splitting via external strain fields allows for control over the coupling between the QD and the resonator mode \cite{Daraei2006Control, Daraei2007Control, Whittaker2007High, Gerhardt2019Polarization}.
To prove this, we performed a study using a pillar with an elliptical cross section (diameter of $\approx$ 2.8~$\upmu$m, H/V splitting of 120~$\upmu$eV without applied bias to the piezoelectric actuator). This pillar embeds a blue detuned QD (detuning of $\Delta \textup{E}_\textup{{X-C}}$ = 370~$\upmu$eV) at a sample temperature of 9~K (spectrum depicted in Fig. \ref{fig:contour_pol_purcell}(a)).  As we increased the sample temperature, we tuned the emitter (X) through the resonance of the cavity (C) (see Fig \ref{fig:contour_pol_purcell} (b)). We observed an enhancement of the emission intensity as an indicator of coupling of the emitter to the cavity. The spectrum taken when the emission line and cavity are in resonance is colored green. At the sample temperature of 18~K, the QD-cavity detuning has the same absolute value as at 9~K, i.e. the QD is red detuned by $\Delta \textup{E}_\textup{{X-C}}$ = -380~$\upmu$eV.

In order to quantify the Purcell enhancement of our system, we plot the intensity of the QD as a function of the emitter-cavity detuning. The result is shown in the inset of Figure \ref{fig:contour_pol_purcell} (b). The data were fitted using the following equation:

\begin{align}
I_{X, cav}(\Delta) \propto \frac{F_P\,L(\Delta)}{1 + F_P\,L(\Delta)} \equiv \beta(\Delta),
\label{eq:I_beta}
\end{align}

where the function $\mathrm{L}(\Delta) = 1/(1+\Delta^2/\mathrm{\kappa^2_0})$ is a Lorentzian of width $\mathrm{\kappa_0}$ describing the empty cavity line shape and $\upbeta$($\Delta$) quantifies the overlap of the exciton emission pattern with the cavity mode \cite{Munsch2009}. The fit indicates a moderate Purcell factor of $\mathrm{F_P}$ = 3.1\,$\pm$\,0.4 confirming the existence of a coupling between QD and cavity, which becomes a necessary assessment for further analysis.

In Fig. \ref{fig:contour_pol_purcell}\,(c) we show the polarization resolved intensity for the two selected detuning values shown in Fig. \ref{fig:contour_pol_purcell}\,(a). We observe that due to the detuning the QD emission acquires a distinct degree of linear polarization, defined as DOLP~=~$\mathrm{(I_H-I_V)/(I_H+I_V)}$) = $\pm\,37~\%$. The emission couples to the cavity polarization mode that is spectrally closer to it. This proves that the QD polarization is strongly influenced by the cavity splitting, in agreement with previous reports on coupled elliptical QD-micropillar cavities \cite{Daraei2006, lee2014polarized, Gerhardt2019Polarization}. \\

\begin{figure*}[bth]
\begin{center}
\includegraphics[width=0.96\textwidth]{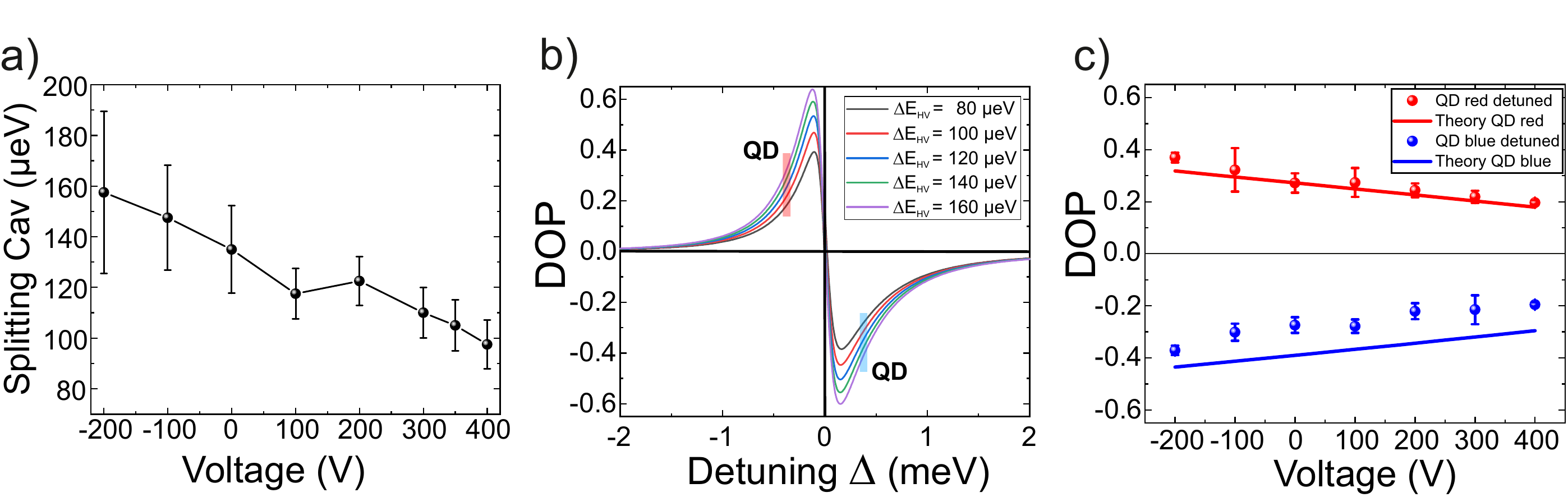}
\caption{(a) The splitting of the fundamental cavity mode as a function of the applied piezo voltage unambiguously shows the trend towards bigger splittings towards voltages. (b) Theoretical plot of the degree of polarization of the cavity modes as a function of the detuning. The different colors are different values of the fundamental cavity mode splitting $\Delta \mathrm{E_{HV}}$ varying from 80 to 160 $\upmu$eV. At any spectral position, the DOLP increases with increasing cavity mode splitting. The red and blue bars depict the detuning ($\mathrm{\Delta_{XC}\approx \pm~380~\upmu eV}$) of the QD in the experiments. In (c), the solid curves depict the theoretical DOLP value at the QD-cavity detuning of $\mathrm{\Delta_{XC} \approx 380~\upmu eV}$ [red and blue colored in (b)] and the measured DOLP values for blue and red detuned QD. While theory and experiment deliver almost identical values for the red detuned case, the blue detuned DOLP reveals a systematical deviation down compared to the theoretical expectation.}
\label{fig:Splitting_DOPs}
\end{center}
\end{figure*}

As discussed earlier, we can modify the cavity anisotropy by inducing strain to the system.  
The associated splitting of the fundamental mode we measured is depicted in Fig.~\ref{fig:Splitting_DOPs} (a), as a function of the applied piezo bias. While at zero voltage a considerable splitting is already present (likely related to pre-stress arising during device processing), we observe a decrease of the polarization splitting towards higher voltages, and further increase of the splitting as we apply a negative voltage. 

Following the approach given by Lee and Lin (2014) \cite{lee2014polarized}, in Fig.~\ref{fig:Splitting_DOPs}(b) we plot the theoretically expected DOLP of an emitter as a function of the detuning $\mathrm{\Delta}$ for various cavity splittings $\mathrm{\Delta E_{HV}}$. Here, we made use of the experimentally determined Purcell factor $\mathrm {F_P}$ as well as the measured linewidth $\mathrm{\gamma_{H,V}}$ of H and V mode, respectively. We marked the dot-cavity detuning of $\mathrm{\Delta_{XC}\approx \pm~380~\upmu eV}$ red and blue. As the theoretical curve indicates, the DOLP of the QD is expected to increase for an increased splitting of the fundamental cavity mode, in particular for the case of moderate emitter-cavity detunings. \\

To prove this experimentally, we have recorded the DOLP as a function of applied strain for both positive and negative detuning. To compensate the strain-induced spectral QD shift, we have re-adjusted the sample temperature in each experiment. 
The result is plotted in Fig. \ref{fig:Splitting_DOPs}(c). We observe an interplay of emitter polarization and applied strain, which in absolute values behaves identically for blue- and red detuned conditions. By increasing the cavity splitting with strain, the polarization increases (-200 V), and reduces towards large positive bias where the H/V eigenmodes of the cavity are almost degenerate. 
To compare the measured DOLP values (red and blue) with theory, we plotted the theoretical values (solid lines) from Fig. \ref{fig:Splitting_DOPs}(b) as a function of the corresponding voltages that have been applied on the piezoelectric actuator in Fig. \ref{fig:Splitting_DOPs}(c). 
While theory and experiment deliver agreement for the red detuned case, the blue detuned DOLP reveals a systematical deviation compared to the theoretical expectation. This might be explained by the modest Purcell enhancement which suggests that our emitter is not centered in the micropillar cavity. This leads to a weaker field strength at its position, and consequently to a smaller effective DOLP. Nevertheless, the overall trend towards a higher DOLP with increasing splitting of the fundamental modes is well confirmed by our measurements.

\section{Conclusion}

In conclusion, we have demonstrated tuning of the fundamental cavity mode polarization splitting in a micropillar cavity by applying an external mechanical stress. The tuning behaviour can be understood as a consequence of anisotropic external strain transmitted to the micropillar acting on its shape as well as on the material's birefringeance. Reconfigurably shaping the ellipticity and birefringeance of a micropillar cavity device is an important step towards achieving the control over the polarization properties of coupled QD-emitter systems, which is of crucial importance for the further improvement of high-performance QD-cavity single photon sources. Here, we demonstrate the first steps by utilizing the polarization anisotropic Purcell enhancement to tune the polarization of a quantum emitter by means of reshaping the cavity mode properties which it couples to. Our findings can be straight-forwardly adapted to other microcavity systems, by instance to tune spin-orbit coupling \cite{Sala2015Spin}, a crucial component in the construction of photonic topological insulators \cite{Klembt2018}. Our findings can further be utilized to impact the pseudo-spin of bosonic condensates of light-matter coupled hybrid systems \cite{Klaas2019Spinselection}, which currently are gaining interest in the construction of solid state quantum bits \cite{Sedov2019}.

\label{results}

The authors would like to thank S. Kuhn for help in sample preparation. We acknowledge funding by the DFG within the project SCHN1376-5.1 and PR1749/1-1. Further, we acknowledge financial support by the State of Bavaria and the German Ministry of Education and Research (BMBF) within the project Q.Link.X (FKZ 16KIS0871). Project HYPER-U-P-S has received funding from the QuantERA ERA-NET Cofund in Quantum Technologies implemented within the European Union's Horizon 2020 Programme. AP would like to thank the Swedish Research Council. J. M.-S. acknowledges finantial support from the Ramón y Cajal Program from the Government of Spain (RYC2018-026196-I) and the ClarínProgramme from the Government of the Principality of Asturias and a Marie Curie-COFUND grant (PA-18-ACB17-29).

\end{document}